\documentclass[11pt,twoside,twocolumn]{article}

\usepackage{amsfonts,amsmath,amssymb,bm,abstract,balance}

\title{\textbf{Formation of bound states of electrons \\
in spherically symmetric oscillations of plasma}}

\author{Maxim Dvornikov\thanks{E-mail: maxim.dvornikov@usm.cl}
\\
\small{Departamento de F\'{i}sica y
Centro-Cient\'{i}fico-Tecnol\'{o}gico de Valpara\'{i}so,} \\
\small{Universidad T\'{e}cnica Federico Santa Mar\'{i}a, Casilla 110-V,} \\
\small{Valpara\'{i}so, Chile and} \\
\small{IZMIRAN, 142190, Troitsk, Moscow region, Russia}}

\date{}

\begin{document}

\twocolumn[\maketitle
\begin{onecolabstract}
We study spherically symmetric oscillations of electrons in plasma
in the frame of classical electrodynamics. Firstly, we analyze the
electromagnetic potentials for the system of radially oscillating
charged particles. Secondly, we consider both free and forced
spherically symmetric oscillations of electrons. Finally, we
discuss the interaction between radially oscillating electrons
through the exchange of ion acoustic waves. It is obtained that
the effective potential of this interaction can be attractive and
can transcend the Debye-H\"{u}ckel potential. We suggest that
oscillating electrons can form bound states at the initial stages
of the spherical plasma structure evolution. The possible
applications of the obtained results for the theory of natural
plasmoids are examined.
\\
\end{onecolabstract}]

\section{Introduction}

Theoretical explanation of the existence of self-sustained
spherically symmetric plasmoids is still a challenge in plasma
physics~\cite{Ste99}. A static bunch of charged particles confined
by its own internal forces does not seem to be stable. The
hydrodynamical pressure $p$ and the magnetic pressure $B^2/8\pi$
will try to expand a plasmoid (see Ref.~\cite{LanLif82p322})
making its existence impossible in the absence of external forces
such as gravity, etc.

However a spherical plasmoid can be implemented in the form of
spherically symmetric oscillations of electrons in
plasma~\cite{DvoDvo07}. This kind of plasma pulsation does not
have a magnetic field (see Sec.~\ref{PHIA} below) and thus does
not lose energy for radiation. This fact can explain the relative
stability of spherical plasmoid generated in natural conditions.

It is, however, known that the frequency of free electrons
oscillations in plasma cannot be less than the Langmuir frequency.
For an electrons density in plasma of $\sim
10^{15}\thinspace\text{cm}^{-3}$ the Langmuir frequency is about
$100\thinspace\text{GHz}$. It is rather difficult to create a
strong external field of such a high frequency in natural
conditions to generate a plasmoid. Therefore one should point out
a physical mechanism that acts at the initial stages of the
plasmoid evolution, during which one can encounter a relatively
low-frequency external field and which makes the plasmoid
generation possible. We suggest that plasma superconductivity can
be one such mechanisms.

The idea that a dense plasma can reveal superconducting properties
was put forward in Ref.~\cite{Dij80Zel08}. The superconductivity
seems to play a key role at the stages of the plasmoid formation.
However it is difficult to find the physical mechanism which would
be responsible for the appearance of the superconducting phase in
rather hot plasma.

It is known that the formation of a bound state of two electrons,
a Cooper pair, underlies the superconductivity phenomenon in
metals. A Cooper pair is formed when the electrostatic repulsive
field of an electron is shielded by the effective attractive
interaction due to the exchange of virtual phonons. Such a bound
state is destroyed when the temperature of a metal exceeds a few
kelvin degrees~\cite{Mad80p315}.

Although the temperature of plasmas in natural conditions is far
greater than a typical temperature of a superconducting metal one
can find physical processes leading to the appearance of the
effective attraction between electrons. A charged test particle,
e.g. an electron, moving in plasma is known to emit ion acoustic
waves. Therefore a test electron can be surrounded by a cloud of
positively charged ions. Under certain conditions this effective
potential can screen the repulsive interaction between two
electrons and result in the creation of a bound state (see
Ref.~\cite{NamAka85}). This phenomenon, as well as the exchange of
dust acoustic waves, can lead to the effective attraction of dust
particles in a dusty plasma~\cite{Shu01}. Note that the formation
of bound states of electrons in plasma due to the exchange of ion
acoustic waves is analogous to the Cooper pairs
formation~\cite{NamVlaShu95}.

In this work we study spherically symmetric plasma structures
using the method of classical electrodynamics. Note that in
Ref.~\cite{DvoDvo07} we considered a spherical plasmoid as quantum
oscillations of electrons in plasma. Firstly, in Sec.~\ref{PHIA},
we examine the electromagnetic potentials for the spherically
symmetric motion of charged particles and find a gauge in which
the vector potential is zero. Secondly, in Sec.~\ref{CLOSC}, we
study free and forced spherically symmetric oscillations of
electrons on the basis of the system of equations of classical
plasma hydrodynamics. In Sec.~\ref{WAKE}, using the methods of
Ref.~\cite{NamAka85} we calculate the scalar potential created by
a test electron participating in forced oscillations in plasma. We
examine the conditions when the effective potential is attractive
and consider the possibility of forming a bound state. Finally, in
Sec.~\ref{DISC}, we examine possible applications of the obtained
results to the description of natural spherical plasmoids.

\section{A description of
spherically symmetric oscillations of electrons in plasma based on
classical electrodynamics\label{CLPL}}

In this section we will study oscillations of elections in plasma
which have the spherical symmetry. The method of classical
electrodynamics will be used to describe this process. Firstly, in
Sec.~\ref{PHIA}, we will be interested in the various choices of
electromagnetic potentials for such a system. Then, in
Sec.~\ref{CLOSC}, we will obtain the exact solution of the
hydrodynamical equations which describes oscillations of electrons
density as well as the dispersion relation for these oscillations.

\subsection{Electromagnetic potentials in the system of
radially oscillating particles\label{PHIA}}

It is clear that the scalar potential of an electric field
$\varphi$ in the system making spherically symmetric pulsations
can depend only on the radial coordinate, $\varphi(r,t)$. The
vector potential $\mathbf{A}$ has only radial component which is
also a function of only radial coordinate, $A_r(r,t)$. Due to the
existence of the spatial dispersion in plasma, oscillations of
charged particles should vanish on big distances from the center
of the system, i.e. $A_r(r,t) \to 0$ at $r \to \infty$.

Suppose that we have found the potentials $\varphi$ and
$\mathbf{A}$. Now we can make the gauge transformation,
$\mathbf{A}' = \mathbf{A} + \nabla f$ and $\varphi' = \varphi +
(1/c)\partial f/\partial t$, where $f(r,t)=\int_r^\infty
A_r(r',t)\mathrm{d}r'$. This gauge transformation does not change
the electric field. Note that the magnetic field is identically
equal to zero for a spherically symmetric motion of charged
particles, $\mathbf{B}= \nabla \times \mathbf{A} = 0$. For the
chosen function $f$ we obtain that the vector potential can be
eliminated in all the space. Now the electric field has the form,
$\mathbf{E}=-\nabla\varphi'$. In the following we will omit the
prime in the definition of the scalar potential.

\subsection{Classical plasma hydrodynamics description
of a spherical plasmoid\label{CLOSC}}

In the first approximation we suggest that only electrons
participate in oscillations of plasma since the mobility of ions
is low. In the absence of collisions and other forms of
dissipation the system of the hydrodynamic equations for the
description of plasma oscillations can be presented in the
following way (see Ref.~\cite{Jac65p369}):
\begin{align}
  \notag
  & \frac{\partial n_e}{\partial t} + \nabla \cdot (n_e\mathbf{v}) = 0, \\
  \notag
  & \frac{\partial\mathbf{v}}{\partial t} + (\mathbf{v} \cdot \nabla)\mathbf{v} =
  -\frac{e}{m}\mathbf{E}-\frac{1}{m n_e}\nabla p, \\
  \notag
  & \nabla \cdot \mathbf{E} = -4\pi e(n_e-n_i)+4\pi\rho_\mathrm{ext}(\mathbf{r},t), \\
  \label{hydrodyn}
  & \frac{\partial\mathbf{E}}{\partial t} =
  4\pi e n_e \mathbf{v},
\end{align}
where $n_e$ is the electrons density, $n_i$ is the ions density,
$p$ is the plasma pressure, $m$ is the mass of the electron and
$e>0$ is the proton charge. In Eq.~\eqref{hydrodyn} we include the
possible external source $\rho_\mathrm{ext}$ and take into account
that the magnetic field is equal to zero (see Sec.~\ref{PHIA}).

Supposing that $n_i = n_0$ and the deviation of the electrons
density from the equilibrium value is small, $n_e-n_0 = n \ll
n_0$, we can linearize the system~\eqref{hydrodyn} and obtain the
single differential equation for the perturbation of the electrons
density,
\begin{equation}\label{closcil}
  \frac{\partial^2 n}{\partial t^2}+\omega_e^2 n -
  \frac{1}{m}
  \left(
    \frac{\partial p}{\partial n}
  \right)_0 \nabla^2 n = \frac{4\pi e n_0}{m}\rho_\mathrm{ext},
\end{equation}
where $\omega_e=\sqrt{4\pi e^2 n_0/m}$ is the plasma frequency for
electrons and $(\partial p/\partial n)_0$ is the derivative taken
at $n_e=n_0$. The latter quantity depends on the equation of state
of electrons in plasma.

In the absence of the external source the spherically symmetric
solution of Eq.~\eqref{closcil} has the form
\begin{equation}\label{clsol}
  n(r,t) = A \cos(\omega t)\frac{\sin\gamma r}{r},
\end{equation}
where $A$ is the constant chosen to satisfy the condition $|n| \ll
n_0$. The frequency of oscillations $\omega$ and the length scale
parameter $\gamma$ are related by the following identity:
\begin{equation}\label{cldisprel}
  \omega^2=\omega_e^2+\frac{1}{m}
  \left(
    \frac{\partial p}{\partial n}
  \right)_0\gamma^2,
\end{equation}
which shows that free oscillations with $\omega \geq \omega_e$
exist in plasma. However, if $\rho_\mathrm{ext} \sim \cos\Omega
t$, it is clear that forced oscillations with $\Omega<\omega_e$
can be also excited.

In Ref.~\cite{DvoDvo07} we studied spherically symmetric
oscillations of electrons in plasma using the quantum mechanical
approach. In that paper, we solved the non-linear Schr\"{o}dinger
equation for the wave function normalized on the number density of
electrons, $|\psi(\mathbf{r},t)|^2=n_e(\mathbf{r},t)$. It was
obtained that in the spherically symmetric case the density of
electrons has the form, $n_e(r,t)=n_0+A\cos(\omega t)\sin(\gamma
r)/r+\dotsb$, i.e. is similar to Eq.~\eqref{clsol}. However the
dispersion relation in Ref.~\cite{DvoDvo07} was different from
Eq.~\eqref{cldisprel}. Moreover quantum oscillations of electrons
reveal the typical size of the system, where the most intensive
oscillations happen, $L = \pi/\gamma =
\pi\sqrt{\hbar/2m\omega_e}$, at the critical frequency
$\omega=2\omega_e$. On the contrary, if we use the classical
electrodynamics method, the parameter $\pi/\gamma$, in principle,
can be arbitrary. Of course, Eq.~\eqref{cldisprel} is valid only
for quite long waves, when $\gamma \ll k_e$ (see
Ref.~\cite{Jac65p374}), where $k_e$ is the Debye wave number (see
the definition in Sec.~\ref{WAKE}).

\section{Formation of bound states of electrons at the initial
stages of the spherical plasmoid evolution\label{WAKE}}

Any particle in plasma has a two-fold life. On the one hand it is
a test particle moving through plasma and interacting with the
whole plasma rather than with separate plasma particles. On the
other hand any test particle is a part of plasma and hence it
contributes to self-consistent electromagnetic fields in plasma.
In this section we will use the method of test particles (see
Refs.~\cite{NamAka85,NamVlaShu95}) to calculate the effective
potential of a radially oscillating electron.

Let us study the electric field created by electrons participating
in spherically symmetric motion considering each electron as a
test particle with the charge $q$. Each of the test particles is
taken to interact with the rest of hot electrons, having the
temperature $T$, and with cold ions. The permittivity of this
plasma has the form,
\begin{equation}\label{perm}
  \varepsilon(\mathbf{k},\omega)=
  1+
  \left(
    \frac{k_e}{k}
  \right)^2-
    \left(
    \frac{\omega_i}{\omega+\mathrm{i}0}
  \right)^2,
\end{equation}
where $k_e=\sqrt{4\pi n_0 e^2/T}$ is the Debye wave number and
$\omega_i=\sqrt{4\pi (Z e)^2 n_0/M}$ is the plasma frequency for
ions with the mass $M$ and the charge $Ze$, $Z$ is the degree of
the ionization of an ion.

We suppose that a test particle makes harmonic oscillations around
the point $\mathbf{r}_0$ with the frequency $\Omega$ and the
amplitude $\mathbf{a}$: $\mathbf{r}'(t) = \mathbf{r}_0 +
\mathbf{a}\sin \Omega t$. To find the electric potential created
by one of the charged particles we should account for the Maxwell
equation for the electric displacement field, $\nabla \cdot
\mathbf{D} = 4 \pi q \delta^3(\mathbf{r}-\mathbf{r}')$.
%
%
Expressing the electric field as $\mathbf{E}(\mathbf{k},\omega) =
- \mathrm{i} \mathbf{k} \varphi(\mathbf{k},\omega)$, we obtain the
scalar potential of the system in the form
\begin{align}\label{varphi1}
  \varphi(\mathbf{k},\omega) = &
  \frac{4 \pi q}{k^2 \varepsilon(\mathbf{k},\omega)}
  \notag
  \\
  & \times
  \int \mathrm{d}t
  e^{\mathrm{i} \omega t -
  \mathrm{i} \mathbf{k} \cdot (\mathbf{r}_0 + \mathbf{a}\sin \Omega t)}.
\end{align}
We recall that the vector potential can be taken to be equal to
zero for the spherically symmetric system (see Sec.~\ref{PHIA}).

Let us decompose the exponential factor in the integrand of
Eq.~\eqref{varphi1} using the series of Bessel functions of the
$n$-th order,
\begin{equation}\label{bessel}
  e^{-\mathrm{i} \mathbf{k} \cdot \mathbf{a} \sin \Omega t}=
  \sum_{n=-\infty}^{+\infty}
  J_n(\mathbf{k} \cdot \mathbf{a})e^{-\mathrm{i} n \Omega t}.
\end{equation}
Finally on the basis of Eqs.~\eqref{varphi1} and~\eqref{bessel} we
obtain the scalar potential in the form
\begin{align}\label{varphi2}
  \varphi(\mathbf{r},t) = &
  \frac{q}{2\pi^2}
  \sum_{n=-\infty}^{+\infty}
  \int \mathrm{d}\omega\mathrm{d}^3\mathbf{k}
  e^{-\mathrm{i} \omega t + \mathrm{i} \mathbf{k} \cdot (\mathbf{r}-\mathbf{r}_0)}
  \notag
  \\
  & \times
  \delta(\omega-n\Omega)
  \frac{J_n(\mathbf{k} \cdot \mathbf{a})}
  {k^2\varepsilon(\mathbf{k},\omega)}.
\end{align}

To analyze Eq.~\eqref{varphi1} we present the reciprocal of the
permittivity~\eqref{perm} in the following way:
\begin{equation}\label{iaw}
  \frac{1}{\varepsilon(\mathbf{k},\omega)}=
  \frac{k^2}{k^2+k_e^2}
  \left(
    1+\frac{\omega_a^2}{\omega^2-\omega_a^2}
  \right),
\end{equation}
where $\omega_a=k\omega_i/\sqrt{k^2+k_e^2}$ is the dispersion
relation for ion acoustic waves.

Using Eq.~\eqref{iaw} we present $\varphi$ in Eq.~\eqref{varphi2}
as a sum of two terms, $\varphi = \varphi_D + \varphi_W$, where
\begin{align}\label{debye1}
  \varphi_D(\mathbf{r},t) = &
  \frac{q}{2\pi^2}
  \sum_{n=-\infty}^{+\infty}
  \int \mathrm{d}^3\mathbf{k}
  e^{-\mathrm{i} n \Omega t + \mathrm{i} \mathbf{k} \cdot (\mathbf{r}-\mathbf{r}_0)}
  \notag
  \\
  & \times
  \frac{J_n(\mathbf{k} \cdot \mathbf{a})}{k_e^2+k^2},
\end{align}
is the analog of the Debye-H\"{u}ckel screening potential and
\begin{align}\label{wake1}
  \varphi_W(\mathbf{r},t) = &
  \frac{q}{2\pi^2}
  \sum_{n=-\infty}^{+\infty}
  \int \mathrm{d}^3\mathbf{k}
  e^{-\mathrm{i} n \Omega t + \mathrm{i} \mathbf{k} \cdot (\mathbf{r}-\mathbf{r}_0)}
  \notag
  \\
  & \times
  \frac{\omega_a^2}{(n\Omega)^2-\omega_a^2}
  \frac{J_n(\mathbf{k} \cdot \mathbf{a})}{k_e^2+k^2},
\end{align}
is the wake potential due to the emission of ion acoustic
waves~\cite{NamAka85}.

To study the behaviour of the potentials $\varphi_D$ and
$\varphi_W$ we choose the specific coordinate system with
$\mathbf{r}_0 = 0$ and $\mathbf{a} = a\mathbf{e}_z$, where
$\mathbf{e}_z$ is the unit vector along the $z$-axis. We also
decompose the vectors $\mathbf{k}$ and $\mathbf{r}$ using the
cylindrical coordinates: $\mathbf{k} = (k_\rho,k_z,\phi_k)$ and
$\mathbf{r} = (\rho,z,\phi)$.

Taking into account the value of the following integral:
\begin{equation}
  \int_0^{2\pi}\mathrm{d}\phi_k e^{\mathrm{i} k_\rho \cos(\phi_k-\phi)}
  = 2\pi J_0(k_\rho \rho),
\end{equation}
we rewrite the potential $\varphi_D$ in the form,
\begin{align}\label{debye2}
  \varphi_D(\rho,z,t) = &
  2q
  \sum_{n=0}^{\infty} (-1)^n
  \int_0^\infty k_\rho \mathrm{d}k_\rho
  \notag
  \\
  & \times
  \frac{e^{-|z|\sqrt{k_\rho^2+k_e^2}}J_0(k_\rho \rho)}{\sqrt{k_\rho^2+k_e^2}}
  \notag
  \\
  & \times
  \Big\{
    I_{2n}
    \left(
      a \sqrt{k_\rho^2+k_e^2}
    \right)
    \cos[2n \Omega t]
    \notag
    \\
    & \pm
    I_{2n+1}
    \left(
      a \sqrt{k_\rho^2+k_e^2}
    \right)
    \notag
    \\
    & \times
    \sin[(2n+1) \Omega t]
  \Big\},
\end{align}
where $I_n(x) = (\mathrm{i})^{-n} J_n(\mathrm{i}x)$ is the Bessel
function of the imaginary argument.
In Eq.~\eqref{debye2} the `$+{}$' sign stands for $z>0$ and
`$-{}$' for $z<0$.

The electromagnetic field of a charged linear oscillator in
vacuum, $\varepsilon=1$, was studied in Ref.~\cite{SokTer74}. The
scalar potential found in that book is different from
Eq.~\eqref{debye2} since in Ref.~\cite{SokTer74} the problem of
radiation of a linear oscillator was considered and the Lorentz
gauge for potentials was used. We study the case of the
spherically symmetric motion of plasma. As we demonstrated in
Sec.~\ref{PHIA} such a system does not have any magnetic field and
thus cannot emit radiation. In our situation it is more convenient
to use the gauge in which $\mathbf{A}=0$. That is why we get a
different expression for the scalar potential.

It is interesting to analyze Eq.~\eqref{debye2} in the static
limit. For the test particle at rest, i.e. at $a \to 0$, only the
term with the function $I_0(x)$ survives. It is possible to show
that in this limit Eq.~\eqref{debye2} transforms to
$\varphi_D(\rho,z,t) = (q/r) e^{-k_e r}$,
%
%
where $r=\sqrt{z^2+\rho^2}$. We can see that
one recovers the usual form of the Debye-H\"{u}ckel potential.
This analysis justifies our definition of $\varphi_D$.

Now we study the wake potential~\eqref{wake1} using the same
technique as for $\varphi_D$. We present the value of $\varphi_W$
in the cylindrical coordinates as
\begin{align}\label{wake2}
  \varphi_W(\rho,z,t) & =
  -2q
  \int_0^\infty k_\rho \mathrm{d}k_\rho
  \frac{e^{-|z|\sqrt{k_\rho^2+k_e^2}}J_0(k_\rho \rho)}{\sqrt{k_\rho^2+k_e^2}}
  \notag
  \\
  & \times
  I_0
  \left(
    a \sqrt{k_\rho^2+k_e^2}
  \right)
  \notag
  \\
  & + 2q
  \sum_{n=1}^{\infty}
  \left\{
  \begin{matrix}
      1 \\
      (-1)^n \
  \end{matrix}
  \right\}
  \frac{\omega_i^2}{\omega_i^2-(n\Omega)^2}
  \notag
  \\
  & \times
  \frac{k_n^3}{k_e^2+k_n^2}
  \int_0^1 \mathrm{d}x
  J_0(k_n \rho \sqrt{1-x^2})
  \notag
  \\
  & \times
  J_n(a k_n x)\sin(k_n|z|x - n \Omega t),
\end{align}
where
\begin{equation}\label{kn}
  k_n = k_e \frac{n\Omega}{\sqrt{\omega_i^2-(n\Omega)^2}}.
\end{equation}
The upper multiplier in Eq.~\eqref{wake2} corresponds to $z>0$ and
the lower one -- to $z<0$. Comparing Eq.~\eqref{debye2} and
Eq.~\eqref{wake2} we can see that the terms containing the
integrals of $I_0(x)$ cancel in the sum of the two potentials.

To derive Eq.~\eqref{wake2} we suggest that $n\Omega<\omega_i$,
i.e. the wake potential appears only for non-rapid oscillations.
For example, we can consider the forced oscillations of electrons
in plasma described in Sec.~\ref{CLOSC} and suppose that $\Omega$
is a bit less than $\omega_i$. It means that only the first
harmonic is excited.

Let us study Eq.~\eqref{wake2} at the line of the test particle
oscillations, $\rho = 0$, and at big distances from the test
particle, $|z| \gg a$. Putting $n=1$ we obtain from
Eq.~\eqref{wake2}
\begin{equation}\label{wake3}
  \varphi_W(z,t) \approx \mp
  q \frac{\Omega^2}{\omega_i^2-\Omega^2}
  \frac{k_1 a}{|z|}
  \cos(k_1 |z| - \Omega t),
\end{equation}
where the `$-{}$' sign stands for $z>0$ and `$+{}$' for $z<0$.

Returning to the general Eq.~\eqref{debye2} for $\varphi_D$ we
also rewrite it for $\rho = 0$ as
\begin{align}\label{debye4}
  \varphi_D(z,t) = &
  \frac{2q}{|z|}
  \sum_{n=0}^{\infty} (-1)^n
  \int_{k_e|z|}^\infty \mathrm{d}x
  e^{-x}
  \notag
  \\
  & \times
  \Big\{
    I_{2n}
    \left(
      \frac{a}{|z|}x
    \right)
    \cos[2n \Omega t]
    \notag
    \\
    & \pm
    I_{2n+1}
    \left(
      \frac{a}{|z|}x
    \right)
    \notag
    \\
    & \times
    \sin[(2n+1) \Omega t]
  \Big\}.
\end{align}
Comparing Eqs.~\eqref{wake3} and~~\eqref{debye4} we can see that
at large distances the non-Coulomb wake potential transcends the
Debye-H\"{u}ckel potential. For example, the term with the
function $I_1(x)$ in Eq.~\eqref{debye4} at $|z| \gg a$ and $k_e a
\ll 1$ has the form
\begin{equation}\label{debye5}
  \mp \frac{2qa}{z^2}e^{-k_e |z|}\sin(\Omega t),
\end{equation}
which has much smaller value than the wake potential~\eqref{wake3}
at $|z|>1/k_e$. The terms which contain the functions $I_n(x)$
with $n>1$ will give contributions to $\varphi_D$ smaller than
that in Eq.~\eqref{debye5}.

The wake potential is attractive when $\cos(k_1 |z| - \Omega t)>0$
for $z>0$ and when $\cos(k_1 |z| - \Omega t)<0$ for $z<0$. We
recall that we study the radial pulsation of plasma. Thus the
coordinate $z$ coincides with the radial direction. Suppose that a
test electron attracts another electron which is situated at the
distance $d$ away the center of the system during the certain
period of time, $\Delta t_1 = \pi/\Omega$. During the next half a
period of the wake potential variation, $\Delta t_2 = \pi/\Omega$,
the same test electron will attract a different electron situated
at the same distance $d$ but closer to center of the system. It
means that a test electron can always attract some charged
particles in plasma.

To form a bound state with another electron in plasma the energy
of interaction of a test electron $e\varphi_W$ should be greater
than the total energy of its oscillations, $E_\mathrm{osc} = m a^2
\Omega^2/2$. Suppose that two electrons are at the distance
$d=1/k_e$ and the amplitude of oscillations $a = 0.1d \ll d$.
Studying the plasma consisting of electrons, with temperature $T =
10^3\thinspace\text{K}$ and number density $n_0 =
10^{15}\thinspace\text{cm}^{-3}$, as well as of singly ionized
nitrogen atoms we obtain that the ratio
$|e\varphi_W|/E_\mathrm{osc}>1$ if $\Omega > 10^{-4}\omega_i$.
Taking into account that $\omega_i \sim
10^{10}\thinspace\text{s}^{-1}$ we obtain that the frequency of
the forced oscillations should be in the region
$10^6\thinspace\text{s}^{-1}<\Omega<10^{10}\thinspace\text{s}^{-1}$.
Thus we get that the effective attraction can take place in the
atmospheric plasma for the reasonable frequencies of an external
field.

\section{Discussion\label{DISC}}

We studied electrons oscillations in plasma using classical
electrodynamics. It is found that spherically symmetric
oscillations are possible in the classical case, although the
dispersion relation~\eqref{cldisprel} is different from the
previously found for quantum oscillations~\cite{DvoDvo07}. We
suggest that the radial pulsation of plasma underlies the rare
atmospheric electricity phenomenon called a ball lightning
(BL)~\cite{Ste99}.

Radial oscillations of electrons in plasma described in the
present work are unbounded and occure in all the space. It is the
case when a plasmoid, BL, propagates far away from any external
surface. The situation changes when one considers spherically
symmetric oscillations of plasma in a cavity inside a dielectric
material. This kind of oscillations was studied in
Ref.~\cite{SteYuVla93} in presence of external electric and
magnetic fields. The exact system of non-linear equations
describing oscillating spatial patterns was obtained and analyzed
both analytically and numerically. The solutions obtained in
Ref.~\cite{SteYuVla93} can correspond to BL passing through
microscopic cracks in a dielectric material, e.g. glass. There are
several reports of such events collected in Ref.~\cite{Sta85}. The
problem of the interaction of BL with external materials should be
analyzed in out future works.

As we demonstrated, in order to generate spherically symmetric
oscillations, one should excite them with rather high frequencies,
$\Omega > \omega_e$ in the classical case (or $\Omega > 2\omega_e$
in the quantum case). The electron plasma frequency is
$(10^{12}-10^{13})\thinspace\text{s}^{-1}$ for $n_0 =
(10^{15}-10^{17})\thinspace\text{cm}^{-3}$. Such high frequencies
are very difficult to obtain in any natural conditions.

We showed in Sec.~\ref{CLOSC} that forced oscillations are
possible with frequencies less than $\omega_e$. Of course, forced
oscillations will decay as soon as the external force is switched
off. There should be a mechanism which provides the smooth
transition from the generation regime of a spherical plasmoid with
the external harmonic source having $\Omega<\omega_e$ to a regime
with self-sustained oscillations having $\Omega \sim \omega_e$.

We suggest that this mechanism could be a formation of bound
states of electrons in plasma. The motion of bound states of
electrons can result in the appearance of a superconducting state
of plasma inside a spherical plasmoid. Previously the idea that
the superconductivity can exist in plasma was put forward in
Ref.~\cite{Mei84}. Without the existence of superconducting phase
electrons participating in radial oscillations will lose their
energy very quickly, because of the various friction mechanisms,
and will recombine into the initial neutral gas.

In Sec.~\ref{WAKE} we obtained that a test electron oscillating in
plasma with the frequency $\Omega\lessapprox\omega_i$ would emit
ion acoustic waves. Thus the test electron appears to be
surrounded by a cloud of `phonons' that shield its repulsive
potential. Under some conditions the effective interaction between
the test electron and other electrons in plasma turns out to be
attractive. This process can lead to the formation of bound states
of electrons.

Note that for the first time the role of `phonons', or acoustic
waves, for the description of the stability of a spherical
plasmoid was discussed in Ref.~\cite{VlaYak78}. However, in that
work, the phonons exchange between ions and neutral atoms was
considered in frames of the quantum theory. The applications to
the obtained results to the theory of BL were also studied in
Ref.~\cite{VlaYak78}.

The existence of the superconducting phase inside BL was
previously proposed in Ref.~\cite{Dij80Zel08}. However in these
papers the dense plasma of BL was already supposed to be in the
superconducting phase and the phenomenological consequences of
this phenomenon were described. No physical mechanisms for the
formation of the superconducting state were proposed. In the
present work we suggest that the exchange of ion acoustic waves
created in spherically symmetric oscillations of electrons in
plasma results in the effective attractive potential between
electrons. This mechanism can lead to the formation of bound
states of electrons and possibly to a superconducting state inside
a spherical plasmoid.

After one switches off the low frequency external field, bound
states of electrons will be destroyed. We assume that during the
superconducting stage of the plasmoid evolution a source of
internal energy should appear. It was suggested in
Ref.~\cite{nuclfus} that nuclear reactions can serve as the energy
source of BL. The recombination of charged particles will be
compensated by the processes of ionization and creation of new
charged particles owing to this internal source of energy of BL.

\section*{Acknowledgments}

The work has been supported by the CONICYT (Chile) through
Programa Bicentenario PSD-91-2006. The author is very thankful to
Sergey Dvornikov and Timur Rashba for helpful discussions.

\balance

\end{document}